\documentclass[12pt]{amsart}
\usepackage[]{graphicx}
\usepackage[]{color}
\usepackage{setspace}
\usepackage{wrapfig}
\usepackage{lineno}
\usepackage{graphicx}
\usepackage{amsmath}
\usepackage{amsfonts}
\usepackage{natbib}
\usepackage{hyperref}
\hypersetup{
    colorlinks=true,
    linkcolor=blue,
    filecolor=magenta,
    urlcolor=cyan,
    citecolor=blue,
}

\usepackage{url}
\usepackage{amsmath,amsxtra,amssymb,latexsym,epsfig,amscd,amsthm,fancybox,epsfig}
\usepackage[mathscr]{eucal}
\usepackage{graphicx}

\usepackage{multicol,xcolor}
\usepackage{epsfig} 
\usepackage{epstopdf}
\usepackage{cases}
\usepackage{subfig}
\usepackage{color}
\usepackage{hyperref}
\usepackage{bigints}
\usepackage[symbol]{footmisc}

\newtheorem{thm}{Theorem}[section]

\newtheorem{lm}{Lemma}[section]

\theoremstyle{definition}

\theoremstyle{remark}

\newcommand{\E}{\mathbb{E}}

\renewcommand{\P}{\mathbb{P}}

\numberwithin{equation}{section}

\newcommand{\red}[1]{{\leavevmode\color{black}#1}}
\newcommand{\XX}{\widehat X}

\newcommand{\bed}{\begin{displaymath}}
\newcommand{\eed}{\end{displaymath}}
\newcommand{\bea}{\bed\begin{array}{rl}}
\newcommand{\eea}{\end{array}\eed}

\newcommand{\barray}{\begin{array}{ll}}
\newcommand{\earray}{\end{array}}

\def\bar{\overline}

\def\a.s{\text{\;a.s.\;}}

\newcommand{\cu}{C}  
\usepackage{chngcntr}
\counterwithout{equation}{section}
\begin{document}
\bibliographystyle{plainnat}
\newcommand{\logC}{\gamma}

\title[A demographic SAS-CFF model]{When do factors promoting genetic diversity also promote population persistence? A demographic perspective on Gillespie's SAS-CFF model }

\author{Sebastian J. Schreiber}
\address{Department of Evolution and Ecology and the Center for Population Biology, University of California, Davis, CA 95616}
\singlespace
\begin{abstract}
 Classical \red{stochastic} demography predicts that environmental stochasticity reduces population growth rates and, thereby, can increase extinction risk. In contrast, in a 1978 \emph{Theoretical Population Biology} paper, Gillespie demonstrated with his \red{stochastic additive scale and concave fitness function (SAS-CFF)} model that environmental stochasticity can promote genetic diversity. Extending the SAS-CFF to account for demography, I examine the simultaneous effects of environmental stochasticity on  genetic diversity and population persistence. Explicit expressions for the per-capita growth rates of rare alleles and the population at low-density are derived. Consistent with Gillespie's analysis, if the log-fitness function is concave and allelic responses to the environment are not perfectly correlated, then per-capita growth rates of rare alleles are positive and genetic diversity is maintained in the sense of stochastic persistence i.e. allelic frequencies tend to stay away from zero almost-surely and in probability. Alternatively, if the log-fitness function is convex, then per-capita growth rates of rare alleles are negative and an allele asymptotically fixates with probability one. If the population's low-density, per-capita growth rate is positive, then the population persists in the sense of stochastic persistence, else it goes asymptotically extinct with probability one. In contrast to per-capita growth rates of rare alleles, the population's per-capita growth rate  is a decreasing function of the concavity of the log-fitness function. Moreover, when the log-fitness function is concave, allelic diversity increases the population's  per-capita growth rate  while decreasing the per-capita growth rate of rare alleles; when the log-fitness function is convex,  environmental stochasticity decreases the  per-capita growth rate of rare alleles, but increases the population's per-capita growth rate.  Collectively, these results  (i) highlight how mechanisms promoting population persistence may be at odds with mechanisms promoting genetic diversity, and (ii) provide conditions under which population persistence relies on existing standing genetic variation.
\end{abstract}
\maketitle

\section{Introduction}

Temporal variation in fitness can have opposing effects on population growth and the maintenance of genetic diversity within a population. This temporal variation typically reduces the long-term growth rate of populations and, consequently, can increase extinction risk~\citep{lewontin-cohen-69}. Indeed, as a second order approximation, \citet{lewontin-cohen-69} showed that temporal variation reduces the long-term population growth rate by one-half of the coefficient of variation of the fitness.  \citet{tuljapurkar-82} showed \red{that} a similar approximation applies to structured populations in serially uncorrelated  environments as part of a series of \emph{Theoretical Population Biology} papers on stochastic demography. This reduction can increase extinction risk by shifting long-term growth rates from positive to negative~~\citep{lewontin-cohen-69,mclaughlin-etal-02,lande-etal-03} or by increasing the likelihood of populations falling below critical densities where inbreeding or Allee effects drive the population extinct~\citep{dennis-02,liebhold-bascompte-03,ogrady-etal-06,jbd-14}.

In contrast to this detrimental impact of temporal variation on population persistence, temporal variation in fitness can promote genetic diversity via balancing selection~\citep{dempster-54,haldane-jayakar-63,gillespie-78,gillespie-80,gillespie-turelli-89,ellner-hairston-94,turelli-81,hatfield1997multispecies,turelli-etal-01,hedrick-06,svardal2015general}. In a 1978 \emph{Theoretical Population Biology} paper, \citet{gillespie-78} introduced the \red{stochastic additive scale and concave fitness function (SAS-CFF)} model to identify when this balancing selection occurs. At that time, \citet{turelli-81} called this model and its analysis as ``the most cohesive and elaborate theory to account for protein polymorphisms by balancing selection.'' \citet{gillespie-78}'s SAS-CFF model assumes there is a physiological activity scale to which alleles contribute additively in a manner that varies stochastically with environmental conditions i.e. the stochastic additive scale. Furthermore, fitness increases in a concave fashion with the physiological activity scale i.e. the concave fitness function. When the  contributions of different alleles have the same mean and are partially correlated or uncorrelated, the SAS-CFF model predicts the maintenance of a polymorphism~\citep{gillespie-77,gillespie-78}. This prediction also  holds  when differences in mean fitness contributions are not too great~\citep{gillespie-80,turelli-gillespie-80,turelli-81}. Therefore, even if one allele with a higher mean fitness would asymptotically fixate in a constant environment, environmental stochasticity can mediate coexistence via a population genetics form of Chesson's storage effect~\citep{chesson-warner-81,chesson-82,chesson-85,chesson-94,hatfield1997multispecies} (see discussion in Section~\ref{sect:discussion}).

These opposing effects of environmental stochasticity of population growth and maintaining genetic polymorphisms raise several questions. First, to what extent does the maintenance of genetic polymorphisms due to environmental stochasticity offset its negative impacts on population growth? In particular, when are genetic polymorphisms necessary for population persistence? Second, to what extent do mechanisms promoting quick recovery of rare alleles also enhance population recovery from low densities? For example, to what extent does the concavity of the fitness function, which promotes genetic diversity, also promote faster population growth? 

To address these questions, I introduce a demographic version of the SAS-CFF model with density-dependent \red{and frequency-dependent} population growth. For this model, recent methods~\citep{arXiv-18} are used to provide a mathematically rigorous analysis of when alleles coexist and when the population persists in the sense of stochastic persistence from both the ensemble perspective~\citep{chesson-78,chesson-warner-81,chesson-82} (i.e. the probability of an allele frequency or the population density being low far into the future is small) and the typical population trajectory perspective~\citep{tpb-09,jmb-11} (i.e. the fraction of time an allele spends at low frequencies or the population spends at low densities is low). When the conditions for stochastic persistence are violated for the alleles or the population, respectively, I show that an allele asymptotically fixates with probability one or the population asymptotically goes extinct with probability one. The conditions for stochastic persistence versus asymptotic extinction are determined by the realized per-capita growth rates of alleles or the population when their frequency or density, respectively, are low. These per-capita growth rates not only determine whether the population persists and alleles coexist, but also determine how quickly alleles and the population recover when rare.

\section{Models and Methods}

\subsection{The Model} I model a random mating population of a diploid  species with discrete, non-overlapping generations. The fitness of each individual is determined by a single multiallelic locus and the population density $N$. There are $k$ possible alleles, $A_1,A_2,\dots,A_k$ at the locus of interest. Following Gillespie~\citep{gillespie-78}, these alleles contribute additively  to a physiological activity scale. For genotype $A_iA_j$, the activity level in generation $t$ equals $(Y_i(t)+Y_j(t))/2$ where $Y_i(t),Y_j(t)$ are the additive contributions of alleles $i$ and $j$, respectively. These activity levels are translated into the low-density fitness via an increasing function $\phi$ of the activity scale. Namely, the low-density fitness of an individual with genotype $A_iA_j$ equals
\[
\phi\left(\frac{Y_i(t)+Y_j(t)}{2} \right). 
\] 
This fitness function $\phi$ can be concave, linear, or convex, but needs to be three-times differentiable i.e. a $C^3$ function. Hence, the ``C" in SAS-CFF here corresponds to $\phi$ being $C^3$ not being concave.  The activity levels $Y_i(1),Y_i(2),Y_i(3),\dots$ for each allele $i$ are assumed to \red{be} independent and identically distributed in time. \red{For the analysis, I will also assume that $Y_1(t),\dots,Y_k(t)$ are have the same mean and variance}. 

Unlike \citet{gillespie-78}'s SAS-CFF model, the population density $N(t)$ is not constant across generations $t$. The population is regulated by negative density-dependence which reduces the fitness of an individual by a factor $\red{1-f}(N)$. This density-dependent reduction is severe at high densities   and negligible at low densities i.e. $\lim_{N\to\infty} \red{f}(N)=0$ and $\red{f}(0)=1.$ For example, $\red{f}(N)$ may be given by an over-compensatory model like the Ricker equation $\exp(-aN)$~\citep{ricker-54} or a compensatory model like the Beverton-Holt function $\frac{1}{1+aN}$~\citep{beverton-holt-57}. For simplicity, I assume that density-dependent feedbacks do not differentially impact the genotypes (i.e. density-independent selection)--an assumption that is believed to hold for many natural populations~\citep{prout-80,travis-etal-13}.

To describe the dynamics, let $X_i(t)$ be the frequency of allele $i$ at generation $t$. As there is random mating and non-overlapping generations, the genotypic and allelic frequencies are in Hardy-Weinberg equilibrium at the beginning of each generation i.e. the frequency of genotype $A_iA_j$ equals $2X_i(t)X_j(t)$ for $i\neq j$ and $X_i(t)^2$ for $i=j$. The expected low-density fitness  of a randomly chosen individual with at least one copy of allele $i$ \red{(i.e. the marginal fitness of allele $i$)} equals
\[
W_i(X(t),Y(t))=\sum_{j=1}^kX_j(t)\phi\left(\frac{Y_i(t)+Y_j(t)}{2} \right)
\]
\red{where $X(t)=(X_1(t),X_2(t),\dots,X_k(t))$ is the vector of allelic frequencies and $Y(t)=(Y_1(t),Y_2(t),\dots,Y_k(t))$ is the vector of allelic contributions to the physiological activity.} The expected low-density fitness of a randomly chosen individual in the population is 
\[
W(X(t),Y(t))=\sum_{i=1}^k X_i(t)W_i(X(t),Y(t)).
\]
Thus, the dynamics of the allelic frequencies and the population density are
\begin{equation}\label{eq:model}
\begin{aligned}
N(t+1)=& N(t)W(X(t),Y(t))\red{f}(N(t))\\
X_i(t+1)=& X_i(t) \frac{W_i(X(t),Y(t))}{W(X(t),Y(t))}.
\end{aligned}
\end{equation}
The state space for the frequency dynamics $X(t)=(X_1(t),\dots,X_k(t))$ is the probability simplex $\Delta=\{x\in [0,1]^k:\sum_{i=1}^k x_i=1\}$. The allelic extinction set $\Delta_0=\{x\in \Delta: \prod_{i=1}^k x_i =0\}$ corresponds to one or more alleles missing from the population. The state space for the eco-evolutionary dynamics $(X(t),N(t))$ is $S=\Delta \times [0,\infty)$ where extinction of the population corresponds to the set $S_0=\Delta \times \{0\}$, and extinction of the population or an allele corresponds to the set $S_{0}\cup \{\Delta_0\times [0,\infty)\}$.

For the analysis of equation~\eqref{eq:model}, I consider the case of small temporal fluctuations and  a probabilistic symmetry of allelic contributions to the physiological activity~\citep{gillespie-77,gillespie-78}. Focusing on this case highlights the main differences between the effects of the convexity of the fitness function, allelic diversity $k$, and stochastic fluctuations on maintaining genetic polymorphisms and population persistence. In particular, I assume a diffusion-type scaling where $Y_i(t)=1+\varepsilon^2\mu+\varepsilon\sigma Z_i(t)$ with $\varepsilon>0$ small, $\E[Z_i(t)]=0$, $\rm{Var}[Z_i(t)]=1$, and $\E[Z_i(t)Z_j(t)]=\rho$ for $i\neq j.$  \red{This diffusion scaling ensures that one can study the simultaneous effects of the mean and variance of the physiological activity on fitness when the fluctuations are small~\citep{turelli1977}. These small noise approximations can work surprisingly well for empirically-based models with large environmental fluctuations~\citep[see, e.g.,][]{turelli-etal-01,buckley_ramula2010}. My particular diffusion scaling assumes a probabilistic symmetry where} the mean and variance of the activity levels for every allele are $1+\varepsilon^2\mu$ and $\varepsilon^2\sigma^2$, respectively, and the correlation of these contributions from two distinct alleles is $\rho$\red{. This symmetry implies that the correlation $\rho$ must lie between $-1/(k-1)$ and $1$}. By a change of variables, I can assume that $\red{\phi(1)=}\phi'(1)=1$. Indeed, without this assumption, the approximations given in Lemmas~\ref{lm:ri},\ref{lm:rN1}, and \ref{lm:rN} still hold if one replaces the mean $1+\varepsilon^2\mu$ with $1+\phi'(1)\varepsilon^2\mu$, the standard deviation $\sigma$ with $\phi'(1)\sigma$, and the curvature $\cu=\phi''(1)$ with $\phi''(1)/(\phi'(1))^2$. \red{Under these assumptions, the convexity of the $\log$-fitness function equals $\logC:=\frac{d}{dx}\Big|_{x=1}\log \phi(x)=\frac{\phi''(1)\phi(1)-\phi'(1)^2}{\phi(1)^2}=\cu-1$.}

\subsection{Definitions and Methods} 

The dynamics of equation~\eqref{eq:model} are analyzed using the methods of \citet{arXiv-18}. These methods provide a mathematically rigorous approach to determining when the alleles or the population persist in the sense of stochastic persistence~\citep{chesson-82,chesson-ellner-89,jdea-11,jmb-11}. Specifically, for a choice of the extinction set $\mathcal{E}$ (e.g. $\Delta_0\times [0,\infty)$ or $S_{0}$, or $S_0\cup \{\Delta_0\times [0,\infty)\}$) the dynamics $Z(t)=(X(t),N(t))$ is \emph{$\mathcal{E}$-stochastically persistent in probability} if 
\begin{equation}
\lim_{t\to\infty}\P\left[\rm{dist}(Z(t),\mathcal{E})\le \delta|Z(0)=z \right]\downarrow 0 \mbox{ as }\delta \downarrow 0\mbox{ whenever }z\notin \mathcal{E},
\end{equation}
where $\rm{dist}(z,\mathcal{E})=\min_{z'\in{\mathcal E}}\|z-z'\|$ is the distance between $z$ and the extinction set. In words, if the process starts in a persistent state, then the probability of being arbitrarily close to the extinction set far in the future is arbitrarily small.  Alternatively, $Z(t)$ is \emph{almost-surely $\mathcal{E}$-stochastically persistent} if 
\begin{equation}
\lim_{T\to\infty}\frac{\#\{1\le t\le T:\rm{dist}(Z(t),\mathcal{E})\le \delta\}}{T}\downarrow 0 \mbox{ as }\delta \downarrow 0\mbox{ almost surely whenever }Z(0)\notin \mathcal{E}.
\end{equation} 
In words, if the process starts in a persistent state, then asymptotically the fraction of time it spends arbitrarily close to the extinction set is arbitrarily small. When both forms of $\mathcal{E}$-stochastic persistence are satisfied, $Z(t)$ is called simply \emph{$\mathcal{E}$-stochastically persistent}. When conditions for $\mathcal{E}$-stochastic persistence are not met, the methods of \citet{arXiv-18} are used to show asymptotic extinction with probability one i.e. $\lim_{t\to\infty}\rm{dist}(Z(t),\mathcal{E})=0$ with probability one.

 The methods for verifying stochastic persistence and asymptotic extinction rely on the per-capita growth rates of alleles when rare or the population at low density. If allele $i$ is infinitesimally rare and the dynamics of the other alleles are characterized by an ergodic, stationary solution $\XX(t)$ with $\widehat X_i(t)=0$, then the dynamics of allele $i$'s frequency $X_i(t)$ can be approximated by the solution of the stochastic linear difference equation
 \begin{equation}\label{eq:linearX}
 \widetilde{X}_i(t+1)=  \widetilde{X}_i(t) \frac{W_i(\XX(t),Y(t))}{W(\XX(t),Y(t))}
 \end{equation}
satisfying
 \[
\log\widetilde{X}_i(t)=\log\widetilde{X}_i(0)+\sum_{\tau=0}
 ^{t-1}\log\frac{W_i(\XX(\tau),Y(\tau))}{W(\XX(\tau),Y(\tau))}.
 \]
 As $\log\frac{W_i(\XX(t),Y(t))}{W(\XX(t),Y(t))}$ is an ergodic stationary sequence, the law of large numbers for ergodic stationary sequences implies that with probability one $\frac{1}{t}\log \widetilde {X}_i(t)$ converges to \emph{the realized per-capita growth rate of  allele $i$ } \red{as $t\to\infty$}:
\begin{equation}\label{eq:one}
r_i(\XX)=\E\left[\log \frac{W_i(\XX(t),Y(t))}{W(\XX(t),Y(t))}\right].
\end{equation} \red{The exponentiated quantity $\exp(r_i(\XX))$ corresponds to the geometric mean of allele $i$'s relative fitness~\citep{haldane-jayakar-63}. Thus,} when $r_i(\XX)>0$, the frequency of allele $i$ is predicted to increase at an exponential rate. When $r_i(\XX)<0$, its frequency is predicted to decrease at an exponential rate.

Alternatively, if the population density is low and the allelic dynamics at this low density are characterized by an ergodic stationary solution $\XX(t)$,  then the dynamics of the population's density $N(t)$ can be approximated by the solution $\widetilde{N}(t)$ of 
 \begin{equation}\label{eq:linearN}
 \widetilde{N}(t+1)=  \widetilde{N}(t)W(\XX(t),Y(t))
 \end{equation}
 given by 
 \[
\log\widetilde{N}(t)=\log\widetilde{N}(0)+\sum_{\tau=0}
 ^{t-1}\log W(\XX(\tau),Y(\tau)).
 \]
Thus, with probability one $\frac{1}{t}\log \widetilde {N}(t)$ converges to \emph{the realized per-capita growth rate of  the population} \red{as $t\to\infty$}: 
\begin{equation}\label{eq:two}
r_N(\XX)=\E[\log W(\XX(t),Y(t))].
\end{equation}
\red{The exponentiated quantity $\exp(r_N(\XX))$ corresponds to the geometric mean of the population's low-density fitness~\citep{lewontin-cohen-69}. Thus,}
when $r_N(\XX)>0$, the population is predicted to increase at an exponential rate.

\section{Results}

\subsection{The maintenance or loss of genetic diversity} I first present sufficient and necessary conditions for $\Delta_0$-stochastic persistence of the allelic dynamics. These results complement the work of \citet{gillespie-77,gillespie-78} in that they apply directly to the discrete-time allelic dynamics $X(t)$, provide mathematically rigorous results for stochastic persistence, and characterize the dynamics (asymptotic fixation) when the persistence condition is violated. In contrast, \citet{gillespie-77,gillespie-78} worked with \red{a} limiting stochastic differential equation and provided sufficient conditions for the existence of a positive stationary distribution.  

Let $\XX(t)$ be an ergodic solution for the allelic dynamics supporting a subset of $\ell<k$ alleles. For any allele $i$ not supported by $\XX$, the following lemma provides an approximation of its per-capita growth rate when rare. This approximation \red{assumes rescaling the fitness function such that $1=\phi(1)=\phi'(1)$ in which case $\logC=\frac{d^2}{dx^2}\Big|_{x=1}\log\phi(x)$ equals the convexity of the $\log$-fitness} function. A proof of this approximation is provided in \ref{appendixA}.

\begin{lm}\label{lm:ri} Let $\XX(t)$ be an ergodic solution for the allelic dynamics such that $\XX_i(t)>0$ if and only if $i\le \ell <k$. For $i>\ell$,
\begin{equation}\label{eq:ri}
r_i(\XX)=\red{-}\varepsilon^2\frac{\sigma^2(1-\rho)\red{\logC}}{4\ell}+O(\varepsilon^3).
\end{equation} 
\end{lm}

Equation~\eqref{eq:ri}  naturally generalizes an expression derived by \citet{gillespie-78} for $k=2$ alleles  to an arbitrary number alleles. It is also equivalent in sign to an expression derived by \citet{turelli-gillespie-80,turelli-81}  for any number of alleles in the diffusion limit. Equation~\eqref{eq:ri} implies that if the allelic responses to the environment are not perfectly correlated (i.e. $\rho<1$) and the log-fitness function is concave (i.e. \red{$\logC<0$}), then the realized per-capita growth rate $r_i(\XX)$ of any missing allele $i$ is positive. As the sign of $r_i(\XX)$ does not depend on the number of alleles $\ell<k$ supported by $\XX$, it follows that $r_i(\XX)>0$ for any allele $i$ not supported by any stationary distribution $\XX$. This positivity ensures stochastic persistence of the alleles. Alternatively, when the log-fitness function is convex  (i.e. \red{$\logC>0$}),  these realized per-capita growth rates are negative. Theorem~\ref{thm:Xpersistence} shows that positive realized per-capita growth rates imply stochastic persistence, while negative values imply asymptotic fixation of an allele with probability one. A proof of Theorem~\ref{thm:Xpersistence} is given in \ref{appendixB}. \red{Recall, the symmetry of the model implies that $\rho$ lies between $-\frac{1}{k-1}$ and $1$.}

\begin{thm}\label{thm:Xpersistence} Assume $\varepsilon>0$ is sufficiently small. If \red{$\logC<0$} and $\red{-\frac{1}{k-1}\le}\rho<1$, then the allelic dynamics $X(t)$ are $\Delta_0$-stochastically persistent. If \red{$\logC>0$} and $\red{-\frac{1}{k-1}\le}\rho<1$, then 
\[
\sum_{i=1}^k \P\left[\lim_{t\to\infty}X_i(t)=1\right]=1
\]
and
\[
\P\left[\lim_{t\to\infty}X_i(t)=1\Big|X(0)=x\right]>0 \mbox{ whenever }x_i>0.
\]

\end{thm}

Beyond determining persistence or extinction, the magnitude of $r_i(\XX)$ determines how quickly allele $i$ increases or decreases when it becomes rare. Equation~\eqref{eq:ri} implies that a rare allele will recover (\red{$\logC<0$}) or be lost (\red{$\logC>0$}) more quickly when there are higher uncorrelated environmental fluctuations (i.e. $\sigma^2(1-\rho)$ is more positive), and there are fewer alleles in the population (i.e. $\ell$ is smaller).

\subsection{Population persistence and growth.} The condition for stochastic persistence of the population depends on whether the frequency dynamics only support a single allele (\red{$\logC>0$}) or a protected polymorphism of all the alleles (\red{$\logC<0$}). When only a single allele is supported by the frequency dynamics (\red{$\logC>0$}), Lemma~\ref{lm:rN1} characterizes the realized per-capita growth rate of the population and, hence via Theorem~\ref{thm:Npersistence1}, the persistence and extinction of the population. Proofs are given in \ref{appendixA} and \ref{appendixB}.

\begin{lm}\label{lm:rN1}
If $\red{\logC>0}$, then 
\begin{equation}\label{eq:rN1}
r_N(\XX)=\varepsilon^2\left( \mu + \frac{\sigma^2\red{\logC}}{2}\right) + O(\varepsilon^3)
\end{equation}
for the ergodic solution $\widehat X(t)=(1,0,\dots,0)$.
\end{lm}

\begin{thm}\label{thm:Npersistence1} Assume $\red{\logC>0}$ and $\varepsilon>0$ is sufficiently small. If $\mu + \frac{\sigma^2\red{\logC}}{2}>0$, then the dynamics $(X(t),N(t))$ are $S_0$-stochastically persistent. If $\mu + \frac{\sigma^2\red{\logC}}{2}<0$, then 
$\lim_{t\to\infty} N(t)=0$ with probability one for all initial conditions. \end{thm}

These results imply that when the log-fitness function is convex ($\red{\logC>0}$), the realized per-capita growth rate \eqref{eq:rN1} increases with environmental variance $\sigma^2$. Hence, the population can persist even if the  fitness at the average activity level ($1+\varepsilon^2\mu$) is less than one i.e. $\mu<0$. Intuitively, \citet{jensen-1906}'s inequality implies that the more convex the fitness function $\phi$, the more positive of an effect of a fixed level of fluctuations in the activity scale on the expected fitness.

When the log-fitness function is concave (i.e $\red{\logC<0}$), the realized per-capita growth rate of the population is given by Lemma~\ref{lm:rN} and, as shown in Theorem~\ref{thm:Npersistence1}, the sign of this realized per-capita growth rate determines stochastic persistence versus asymptotic extinction of the population. Proofs are given in \ref{appendixA} and \ref{appendixB}.

\begin{lm}\label{lm:rN} Assume $\red{\logC<0},\red{-\frac{1}{k-1}\le}\rho<1$ and $\varepsilon>0$ is sufficiently small. Then
\begin{equation}\label{eq:rN}
r_N(\XX)=\varepsilon^2\left(\mu+\red{\frac{\sigma^2\logC}{2}+\frac{\sigma^2\logC(\logC-1)(1-\rho)}{2(1-2\logC)}\left(1-1/k\right)}\right)+O(\varepsilon^3)
\end{equation}
for any ergodic solution $\XX(t)$ with $\XX_i(0)>0$ for all $i$.
\end{lm}

\begin{thm}\label{thm:Npersistence2} Assume $\red{\logC<0},\red{-\frac{1}{k-1}\le}\rho<1$ and $\varepsilon>0$ is sufficiently small. If $r_N(\XX)$ as defined by equation~\eqref{eq:rN} is positive, then equation~\eqref{eq:model} is $S_{0}$-- and $\Delta_0$--stochastically persistent. If equation~\eqref{eq:rN} is negative, then $\lim_{t\to \infty}N(t)=0$ with probability one for all initial conditions. 
\end{thm}

\red{\ref{appendixA} shows that equation~\eqref{eq:rN} is an increasing function of the convexity $\logC$ of the log-fitness function. Hence, contrary to its effects on disrupting allelic diversity, convexity of the log-fitness function lowers the realized per-capita growth rate of the population. \ref{appendixA} also shows that $r_N(\XX)$ is a decreasing function of the environmental variance $\sigma^2$ whenever $\logC<0$ and $-\frac{1}{k-1}\le \rho <1$. Hence, unlike its effects on the growth rate of rare alleles, populations with concave log-fitness function recover more slowly from low densities with greater environmental variation.}

\red{When the fitness function is concave ($\logC<0$), the third term of $r_N(\XX)$ in equation~\eqref{eq:rN} is positive. Due to the factor $(1-\rho)(1-1/k)$ of this third term, this positive effect on the realized per-capita growth rate of the population increases with the number of the number of alleles and decreases with environmental correlation $\rho$. Consequently, for a fixed amount of environmental variation and concavity of the log-fitness function, the realized per-capita growth rate of the population is maximized when $(1-\rho)(1-1/k)=1$. This can occur in two ways: (i) the environmental correlation is maximally negative i.e. $\rho=-\frac{1}{k-1}$, or (ii) there are no correlations and infinitely many alleles i.e. $\rho=1$ and $k\to \infty$.} 

Due $r_N(\XX)$ increasing with the number of alleles (when $\red{\logC<0}$), genetic diversity may be necessary for population persistence. For example, consider the special case where the fitness function is additive ($\phi(x)=x$ in which case $\red{\logC=-1}$) and allelic contributions are independent ($\rho=0$). Then, the population's realized per-capita growth rate satisfies  
\begin{equation}\label{eq:rN2}
r_N(\XX)=\varepsilon^2\left( \mu-\frac{\sigma^2}{6}\left(1+\frac{2}{k}\right)\right)+ O(\varepsilon^3).
\end{equation}
For only one allele ($k=1$), equation~\eqref{eq:rN2} yields the classical approximation $r_N(\XX)\approx \varepsilon^2(\mu-\sigma^2/2)$ for the growth rate of a population in a fluctuating environment \citep{gillespie-73b}. Thus, if the index of dispersion $D:=\sigma^2/\mu$ is greater than $2$ and there is only one allele, then the population tends to extinction with probability one. In contrast, if there are many alleles (i.e. $k\to\infty$), then the population growth rate satisfies $r_N(\XX)\approx \varepsilon^2( \mu -\sigma^2/6)$. Thus, the population persists if the index of dispersion $D$ is less than $6$. Hence, whenever the index of dispersion is between $2$ and $6$, there is a critical number of alleles $\red{k=\frac{2D}{6-D}}$ below which the population would go extinct and above which it would persist.

\begin{figure}
\includegraphics[width=0.75\textwidth]{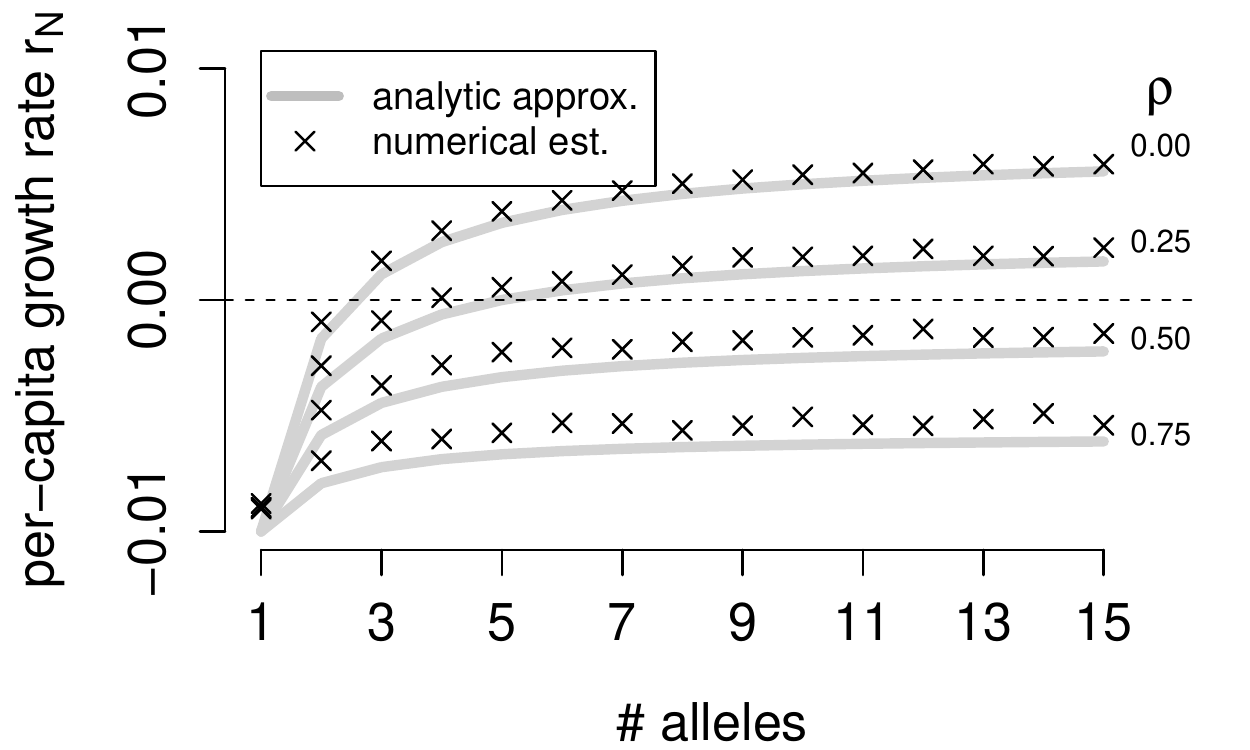}
\caption{The impact of allelic diversity $k$ and correlations $\rho$ on the realized per-capita growth rate $r_N(\XX)$ of the population. For positive $r_N(\XX)$ values the population persists, while for negative values, it goes asymptotically extinct with probability one (cf. Theorem~\ref{thm:Npersistence2}). Numerical estimates (\red{blue} crosses) of $r_N(\XX)$ made by running the allelic dynamics for $1,000,000$ generations and evaluating equation~\eqref{eq:two}; analytic approximation (solid gray) from equation~\eqref{eq:rN}. Model details: $Y_i$ log-normally distributed with mean $1+\varepsilon^2\mu=0.015$ and variance $\varepsilon^2\sigma^2=0.05$; a linear fitness function $\phi(x)=x.$ }\label{fig:A}
\end{figure}

\begin{figure}
\includegraphics[width=0.75\textwidth]{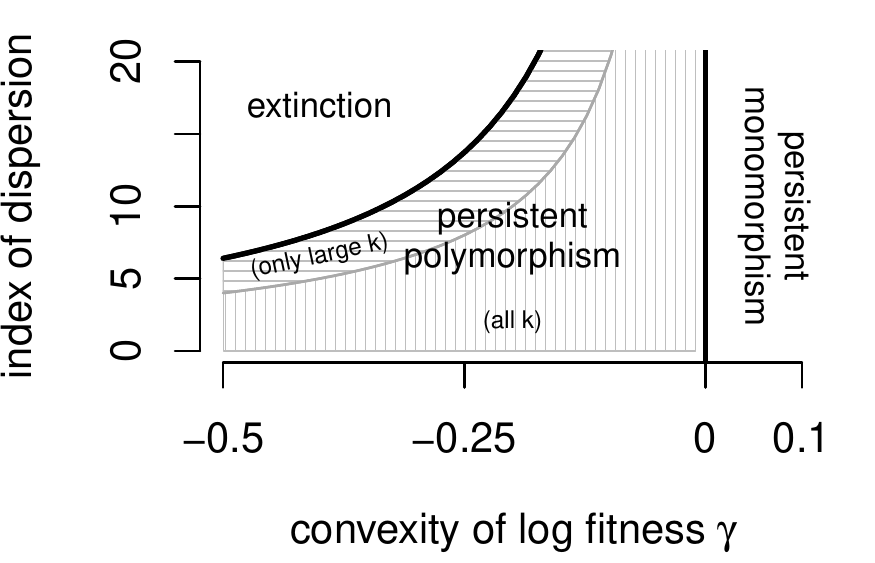}
\caption{\red{The effect of the convexity $\logC$ of the log-fitness function  and the index of dispersion ($D=\mu/\sigma^2$ with $\mu>0$) of environmental stochasticity on population persistence and maintenance of allelic polymorphisms.  Asymptotic extinction always in the top left region. The population persists and can maintain an allelic polymorphism in the shaded region. Population persistence requires sufficiently many alleles ($k$ large) in the horizontally shaded region, and occurs for any number of alleles ($k\ge 1$) in the vertically shaded region. When the log-fitness function is convex ($\gamma>0$), the population persists but is monomorphic.  }  }\label{fig:B}
\end{figure}

Figure~\ref{fig:A} illustrates this surprising phenomen\red{on}. It also illustrates, as can be shown analytically, \red{that} correlated responses of the alleles to the environment make this phenomena less likely to occur: increased allelic diversity has a smaller effect on $r_N(\XX)$ when allelic responses are positive correlated. More generally, whether or not this allelic rescue occurs depends on the index of dispersion $D$ and the convexity $\red{\logC}$ of the log-fitness function $\log \phi$. As illustrated in Figure~\ref{fig:B}, the less concave the log-fitness function is, the broader the range of $D$ value for which genetic diversity is required for population persistence.

\section*{Discussion}\label{sect:discussion}

I began with the question\red{: T}o what extent do factors promoting genetic diversity also promote population persistence? The analysis of a demographic extension of \citet{gillespie-78}'s SAS-CFF model suggests that several factors promoting one, hinder the other. First and foremost, convexity of the log-fitness function $\log \phi$ always promotes population persistence by increasing the population's low-density, per-capita growth rate $r_N(\XX)$, but inhibits genetic diversity by decreasing the per-capita growth rates $r_i(\XX)$ of rare alleles. This difference stems from population growth being determined by the absolute log-fitness and allelic growth rates being determined by their relative log-fitnesses. A convex log-fitness function $\phi$ ($\red{\logC>0}$) coupled with environmental stochasticity increases the expected log-fitness due to \citet{jensen-1906}'s inequality; this increases  $r_N(\XX)$. In contrast, as rare alleles are mostly found in heterozygote individuals while common alleles are also found homozygous individuals, the rare alleles experience a lower variability in its physiological scale (i.e. $\rm{Var}[(Y_i+Y_j)/2]=\sigma^2(1+\rho)/2$ for $i\neq j$) than the common alleles (i.e. $\rm{Var}[(Y_i+Y_j)/2]=\sigma^2$ for $i=j$). Thus, \citet{jensen-1906}'s inequality implies the expected log-fitness of the common allele benefits more from convexity of the log-fitness function $\phi$ than the rare allele. 

For the reasons just outlined, the effects of  environmental stochasticity on genetic diversity and population persistence depend on the convexity of the log-fitness function $\log \phi$. When the log-fitness function is concave ($\red{\logC<0}$), environmental stochasticity promotes protected polymorphisms but has a negative impact on population growth, a conclusion consistent with classical stochastic demography~\citep{lewontin-cohen-69,tuljapurkar-82}.  Alternatively, when the log-fitness function is convex, environmental stochasticity results in alleles being lost faster i.e. $r_i(\XX)$ being more negative. In contrast, environmental stochasticity filtered through a convex log-fitness function promotes population persistence by increasing its low-density per-capita growth rate. Thus, whether the log-fitness function is convex or concave, environmental stochasticity has opposing effects on the  per-capita growth rates of rare alleles and the population at low densities. 

Equations~\eqref{eq:ri} and \eqref{eq:rN}, also highlight that the number of common alleles \red{have} opposing effects on the rate of growth of rare alleles and the population. The positive effect on the population growth rate occurs due to the population-level variance in the physiological activity decreasing with the number of alleles. This effect is similar to spatial bet-hedging in which living in different, partially correlated habitats increases the realized per-capita growth rate of the population~\citep{jansen-yoshimura-98,bascompte-etal-02,jmb-13,amnat-12,jmb-15}. 
In contrast, as the number of common alleles increases, rare alleles have fewer temporal niches to exploit and, consequently, have a lower realized per-capita growth rate. 

In light of Peter Chesson's work on the storage effect~\citep{chesson-warner-81,chesson-82,chesson-85,chesson-94,chesson-03,hatfield-chesson-89,kuang-chesson-10,stump-chesson-17}, it is natural to ask whether \citet{gillespie-78}'s SAS-CFF exhibits the storage effect whose ingredients are~\citep{chesson-94}: (i) species-specific responses to the environment, (ii) positive covariance between environment and competition, and (iii) buffered population growth. Verifying these three ingredients of the storage effects requires expressing the realized per-capita growth $r_i(\XX)$ in terms of two variables for allele $i$: an environmentally dependent variable $E_i(t)$ which doesn't depend on the frequencies of the alleles, and a competition variable $C_i(t)$ that measures the total competition experienced by allele $i$. \red{Consistent with Chesson's theory, one wants to choose $E_i(t)$ such that it has a positive effect on $r_i$, while $C_i(t)$ has a negative effect. One such choice is  $E_i(t)=Y_i(t)$ and $C_i(t)=W(X(t),Y(t))$. For the linear fitness function $\phi(x)=x$, this choice yields}
\[
C_i(t)=W(X(t),Y(t))=\sum_{j,k}X_j(t)X_k(t)\frac{Y_j(t)+Y_k(t)}{2}=\sum_j X_j(t)Y_j(t)\]
 and 
\[
W_i(X(t),Y(t))=\sum_j X_j\frac{Y_i(t)+Y_j(t)}{2}=\frac{E_i(t)}{2}+\frac{C_i(t)}{2}.
\]
Thus, for the linear fitness function $\phi(x)=x$
\[
r_i(X(t))=\E[\log(E_i(t)/C_i(t)+1)]-\log 2.
\]
Ingredient (i) of the storage effects occurs whenever $E_i(t)$ and $E_j(t)$ for $i\neq j$ are not perfectly correlated i.e. $\rho<1$. For ingredient (ii) of the storage effect, the covariance between $E_i(t)$ and $C_i(t)$ equals (assuming $Y_i(t)$ are independent and identically distributed with mean $\bar{Y}$)
\begin{eqnarray*}
\rm{Cov}[E_i(t),C_i(t)|X(t)=x]&=&\E[Y_i(t)-\bar{Y},\sum_j x_j Y_j(t)-\bar{Y}]\\
&=&\E[Y_i(t)-\bar{Y},x_i Y_i(t)-\bar{Y}]=x_i\rm{Var}[Y_i(t)]
\end{eqnarray*}
which is positive whenever $x_i>0$. Finally, ingredient (iii)  requires that the mixed partial derivative of $r_i$ with respect to $E_i$ and $C_i$ be negative. \red{Namely, when the environment favors allele $i$ less, allele $i$ is less effected by competition i.e. decreasing $E_i$ has less a negative effect effect when $C_i$ increases.} This final ingredient is satisfied as  
\[
\frac{\partial^2r_i}{\partial E_i\partial C_i}=\E\left[-\frac{1}{(E_i(t)+C_i(t))^2}\right]<0.
\]
Biologically, this buffering occurs via  alleles residing in heterozygous individuals. Namely,  when environmental conditions are poor for allele $i$ (i.e. $Y_i(t)$ is small), environmental conditions for another allele $j$ are likely to be better. Thus, by residing in heterozygotes $A_iA_j$, allele $i$  gets buffered from these poorer environmental conditions. Verifying the storage effect for the linear fitness function is straight-forward as the SAS-CFF model is equivalent to a special case of \citet{chesson-warner-81}'s lottery model as observed by ~\citet{hatfield1997multispecies} in their diffusion analysis of the multispecies lottery model.
In the case of a nonlinear fitness function $\phi$ verifying the storage effect  appears not to be straightforward as $r_i(X(t))=\E[\log \frac{W_i(X(t),Y(t))}{W(X(t),Y(t))}]$ can not be expressed in terms of $E_i(t)=Y_i(t)$ and $C_i(t)=W(X(t),Y(t))$. Thus, one is left with the challenge of extending \citet{chesson-94}'s framework to general diploid models. 

Other future challenges include developing conditions for population persistence under less restrictive assumptions. For example, \citet{turelli-81} developed results for protected polymorphisms when the mean fitness contributions $\mu_i$ and covariances $\rho_{ij}\sigma_i\sigma_j$ are variable. For diffusive scalings of these parameters,  one can develop expressions for $r_i(\XX)$ (a rescaling of the expression found by \citet{turelli-81}) and $r_N(\XX)$. Using such expressions, one could evaluate the robustness of my main conclusions to asymmetries in the means and covariances. To extend the results beyond the case of a  diffusion scaling is a bigger challenge. While it is possible to use the results of \citet{arXiv-18} to get abstract conditions for stochastic persistence,  the main challenge is getting explicit expressions for these realized per-capita growth rates or identifying their sign under appropriate assumptions. For example, one could ask if $\frac{d^2}{dx^2}\log \phi (x)=\logC$ for all $x$, $\mu_i=\mu$ for all $i$, and $Y_1(t),\dots,Y_k(t)$ are independent and identically distributed, then does the sign of $\logC$ still determine $\Delta_0$-stochastic persistence? One could also include density-dependent selection into the model~\citep{travis-etal-13} and, thereby, make the realized per-capita growth rates $r_i$ depend on both $X(t)$ and $N(t)$. Finally, one could try to relax the assumption of the additive contributions $Y_i(t)$ to the physiological scale by allowing the fitness function to be a nonlinear bivariate function $\phi(Y_i,Y_j).$ I can only hope that the answers to some of these and other challenges will find their way into future issues of \emph{Theoretical Population Biology.}\vskip 0.1in

\textbf{Acknowledgments.} I thank: Peter Chesson for discussions about the relationship between the storage effect and the SAS-CFF model and suggesting to focus on the case of the linear fitness function $\phi(x)=x$; Michael Turelli and Vince Buffalo for several discussions about the SAS-CFF model; William Cuello for carefully reading over the derivations in the Appendices; \red{Two anonymous reviewers,} Matthew Osmond,  Michael Culshaw-Maurer, Sam Fleisher, Kelsey Lyberger, and Dale Clement for providing \red{critical} feedback on earlier stages of this work; The U.S. National Science Foundation for helping fund this work through Grant DMS-1716803.

\bibliography{maladaptive}
\setcounter{section}{0}
\renewcommand*{\thesection}{Appendix \Alph{section}}

\section{Proofs of Lemmas~\ref{lm:ri},\ref{lm:rN1}, and \ref{lm:rN}}\label{appendixA}

Assume that $\phi$ is three times differentiable at $1$, $\phi(1)=1=\phi'(1)$ and $\cu=\phi''(1).$ Also assume that $Y_i(t)=1+\varepsilon^2 \mu+ \varepsilon \sigma Z_i(t)$ where $\E[Z_i(t)]=0$, $\rm{Var}[Z_i(t)]=1$, and $\E[Z_i(t)Z_j(t)]=\rho$ for $i\neq j$. Taylor's theorem implies 
\begin{equation}\label{a}
\phi(1+\varepsilon^2\mu+\varepsilon\sigma(z_i+z_j)/2)=
1+\varepsilon^2\mu+\varepsilon\sigma(z_i+z_j)/2+\frac{\cu\varepsilon^2\sigma^2}{8}(z_i+z_j)^2+O(\varepsilon^\red{3}).
\end{equation}{}
We will use \eqref{a} to prove Lemmas~\ref{lm:ri},~\ref{lm:rN1}, and \ref{lm:rN}.
 
\subsection*{Proof of Lemma~\ref{lm:ri}} 
Let $\XX$ be a stationary distribution supporting $\ell\le k$ alleles. By permuting indices, without loss of generality $\P[\XX_i>0]=1$ for $1\le i\le \ell$ and $\P[\XX_i=0]=1$ for $i>\ell.$ For any $i$, 
\begin{equation}\label{b}
\begin{aligned}
r_i(\XX)&=
\overbrace{\E\left[\log \sum_{j=1}^\ell \XX_j \phi\left(1+\varepsilon^2\mu+\sigma\varepsilon\frac{Z_i+Z_j}{2}\right)\right]}^{\clubsuit}\\
&-\underbrace{\E\left[\log \sum_{1\le r,s\le \ell} \XX_r\XX_s \phi\left(1+\varepsilon^2\mu+\varepsilon\sigma\frac{Z_r+Z_s}{2}\right)\right]}_{\spadesuit}
\end{aligned}
\end{equation}
where $Z_j$ are independent of $\XX$ and have the same distribution as $Z_j(1).$
Approximating $\clubsuit$ in equation~\eqref{b} using $\log(1+x)= x-x^2/2+O(x^3)$, equation~\eqref{a}, $\E[Z_i]=0$, $\E[\XX_j]=\frac{1}{\ell}$ for $j\le \ell$, $\sum_{j=1}^\ell \XX_j=1$, and independence of $\XX_i$ and $Z_j$, one gets 
\begin{eqnarray*}
\clubsuit&=&\varepsilon^2\mu+\frac{\cu\varepsilon^2\sigma^2}{8}\E\left[\sum_{j=1}^\ell \XX_j(Z_i+Z_j)^2\right]-\frac{\varepsilon^2\sigma^2}{8}\E\left[\left(\sum_{j=1}^\ell\XX_j(Z_i+Z_j)\right)^2\right]+O(\varepsilon^3)\\
&=&\varepsilon^2\mu+\frac{\cu\varepsilon^2\sigma^2}{8\ell}\sum_{j=1}^\ell\E[(Z_i+Z_j)^2]-\frac{\varepsilon^2\sigma^2}{8}\E\left[\left(Z_i+\sum_{j=1}^\ell \XX_jZ_j\right)^2\right]+O(\varepsilon^3).
\end{eqnarray*}
As $\E[Z_i^2]=1$ and $\E[Z_iZ_j]=\rho$ for $i\neq j$, one has 
\begin{eqnarray*}
\frac{1}{\ell}\sum_{j=1}^\ell \E[(Z_i+Z_j)^2]&=& \E[Z_i^2]+ \frac{2}{\ell} \sum_{j  =1}^\ell \E[Z_iZ_j]+ \frac{1}{\ell}\sum_{j=1}^\ell \E[Z_j^2]\\
&=&2+2 \left\{ \begin{array}{l}\rho+\frac{1-\rho}{\ell}\mbox{ if }i\le \ell \\
\rho\mbox{ else}\end{array}\right.
\end{eqnarray*}
By defining \[\heartsuit=\sum_{j=1}^\ell \XX_jZ_j,\]  one has 
\begin{eqnarray*}
\E\left[\left(Z_i+\sum_{j=1}^\ell\XX_jZ_j\right)^2\right]&=&\E[Z_i^2]+\frac{2}{\ell}\sum_{j=1}^\ell\E[Z_i Z_j]+\E[\heartsuit^2]\\
&=&1+\E[\heartsuit^2]+2\left\{
\begin{array}{ll}
\rho+\frac{1-\rho}{\ell}&\mbox{ if }i\le \ell\\
\rho&\mbox{ else.}
\end{array}\right.
\end{eqnarray*}
Thus, 
\begin{equation}\label{c}
\clubsuit=O(\varepsilon^3)+\varepsilon^2\mu+\frac{\varepsilon^2\sigma^2}{8}\left\{\begin{array}{ll}
2\cu\left(1+\rho+\frac{1-\rho}{\ell}\right)
-\left(1+2(\rho+\frac{1-\rho}{\ell})+\E[\heartsuit^2]\right)&\mbox{ if }i\le \ell\\
2\cu\left(1+\rho\right)-\left(1+2\rho+\E[\heartsuit^2]\right)&\mbox{ else.}
\end{array}\right.
\end{equation}

To solve for $r_i(\XX)$, \red{assertion (iii) of Proposition 1} from~\citet{arXiv-18} implies that $r_i(\XX)=0$ for all $i\le \ell$. \red{Recall, by assumption} $\ell<k$. Let $\clubsuit^+$ be the value of $\clubsuit$ for an $i\le \ell$ and $\clubsuit^-$ be its value for an $i>\ell$. As $r_i(\XX)=0$ for $i\le \ell$, one has $\clubsuit^+=\spadesuit$. For $i>\ell$, 
\begin{eqnarray*}
r_i(\XX)&=&\clubsuit^--\spadesuit=\clubsuit^--\clubsuit^+\\
&=&\frac{\sigma^2\varepsilon^2}{8}\left(2\cu\left(1+\rho\right)-\left(1+2\rho+\E[\heartsuit^2]\right)\right.\\&&\left.- 2\cu\left(1+\rho+\frac{1-\rho}{\ell}\right)
-\left(1+2(\rho+\frac{1-\rho}{\ell})+\E[\heartsuit^2]\right)  \right)+O(\varepsilon^3)\\
&=&\frac{\sigma^2\varepsilon^2}{4\ell}(1- \cu)(1-\rho) +O(\varepsilon^3).
\end{eqnarray*}
\red{As $\logC=C-1$, this completes the proof of the lemma.}

\subsection*{Proofs of Lemmas~\ref{lm:rN1},\ref{lm:rN}} \red{As  the expression for $r_N(\XX)$ in Lemma~\ref{lm:rN1} corresponds to the corresponding expression in Lemma~\ref{lm:rN} for $k=1$}, it suffices to prove Lemma~\ref{lm:rN}. To this end, assume that $\XX$ is a stationary distribution supporting all alleles. Recall that $r_N(\XX)$ equals  $\spadesuit$ in equation~\eqref{b}, and $\heartsuit=\sum_{j=1}^\ell \XX_jZ_j$. Using $\log(1+x)= x-x^2/2+O(x^3)$, equation \eqref{a}, $\E[Z_j]=0$, and independence of $\XX_i$ and $Z_j$, one has 
\begin{eqnarray*}
\spadesuit&=&\varepsilon^2\mu+\frac{\cu\varepsilon^2\sigma^2}{8}\E\left[\sum_{1\le r,s\le \ell} \XX_r\XX_s(Z_r+Z_s)^2\right]-\frac{\varepsilon^2\sigma^2}{8}\E\left[\left(\sum_{1\le r,s\le \ell} \XX_r\XX_s(Z_r+Z_s)\right)^2\right]+O(\varepsilon^3)\\
&=&\varepsilon^2\mu+\frac{\cu\varepsilon^2\sigma^2}{8}\E\left[2\sum_{r=1}^\ell \XX_rZ_r^2+2\heartsuit^2\right]-\frac{\varepsilon^2\sigma^2}{8}\E\left[\left(2\heartsuit\right)^2\right]+O(\varepsilon^3).\\
\end{eqnarray*}
As $\E[Z_r^2]=1$, $\E[\XX_r]=1/\ell$ and $\XX_r$ is independent of $Z_r$, one has 
\begin{equation}\label{d}
\spadesuit= \varepsilon^2\mu+\frac{\varepsilon^2\sigma^2}{4}\left(\cu+\left(\cu-2\right)\E\left[\heartsuit^2\right]\right)+O(\varepsilon^3).
\end{equation}

Let $\clubsuit^+$ be $\clubsuit$ in equation~\eqref{c} when $i=\ell=k$. As $r_i(\XX)=0$,  $\clubsuit^+=\spadesuit$ and equations \eqref{c}--\eqref{d} can be used to solve for $\E[\heartsuit^2]$ \red{up to} order $\red{\varepsilon}$:
\begin{eqnarray*}
2\cu\left(1+\rho+\frac{1-\rho}{k}\right)
-\left(1+2\left(\rho+\frac{1-\rho}{k}\right)+\E[\heartsuit^2]\right)&=&2\left(\cu+\left(\cu-2\right)\E\left[\heartsuit^2\right]\right)+O(\red{\varepsilon})\\
2\cu\left(\rho+\frac{1-\rho}{k}\right)
-\left(1+2\left(\rho+\frac{1-\rho}{k}\right)\right)&=&\left(2\cu-3\right)\E\left[\heartsuit^2\right]+O(\red{\varepsilon})\\
\frac{2\cu\left(\rho+\frac{1-\rho}{k}\right)
-\left(1+2\left(\rho+\frac{1-\rho}{k}\right)\right)}{2\cu-3}&=&\E[\heartsuit^2]+O(\red{\varepsilon}).
\end{eqnarray*}
Thus,
\[
\begin{aligned}
r_N(\XX)=\spadesuit=& \varepsilon^2\mu+\frac{\varepsilon^2\sigma^2}{4}\left(\cu+\left(\cu-2\right)\frac{2\cu(\rho+\frac{1-\rho}{k})
-\left(1+2(\rho+\frac{1-\rho}{k})\right)}{2\cu-3}\right)+O(\varepsilon^3)\\
=& 
\red{\varepsilon^2\mu+\frac{\varepsilon^2\sigma^2}{4}\left(\cu+\left(\cu-2\right)\frac{(2\cu-2)((1-1/k)\rho+1/k)-1}{2\cu-3}\right)+O(\varepsilon^3)}\\
=& \red{\varepsilon^2\mu+\frac{\varepsilon^2\sigma^2}{4}\left(\cu+\left(\cu-2\right)\frac{(2\cu-2)((1-1/k)(\rho-1)+1)-1}{2\cu-3}\right)+O(\varepsilon^3)}\\
=&\red{\varepsilon^2\mu+\frac{\varepsilon^2\sigma^2}{4}\left(\frac{4\cu^2-10\cu+6}{2\cu-3}+\left(\cu-2\right)\frac{2(\cu-1)(1-1/k)(\rho-1)}{2\cu-3}\right)+O(\varepsilon^3)}\\
=&\red{\varepsilon^2\mu+\frac{\varepsilon^2\sigma^2}{2}\left(\cu-1+\left(\cu-2\right)\frac{(\cu-1)(1-1/k)(\rho-1)}{2\cu-3}\right)+O(\varepsilon^3)}\\
=&\red{\varepsilon^2\mu+\frac{\varepsilon^2\sigma^2}{2}\left(\logC+\frac{\logC \left(\logC-1\right)(1-1/k)(1-\rho)}{1-2\logC}\right)+O(\varepsilon^3)}
\end{aligned}
\]
\red{where the last line follows from $\logC=\cu-1$.}

\subsection*{Properties of the realized per-capita growth rates.}

To understand how the realized per-capita growth rate of the population $r_N$ depends on the convexity \red{$\logC$ of the log-fitness function, define $\alpha=(1-\rho)(1-1/k)$ and 
\[
R=\mu+\frac{\sigma^2\logC}{2}+\frac{\sigma^2\logC(\logC-1)}{2(1-2\logC)}\alpha
\]
which corresponds to the coefficient of the order $\varepsilon^2$ term of $r_N(\XX)$. Taking the first order derivative of $R$ with respect to $\logC$ yields
\[
\frac{\partial R}{\partial \logC}=\frac{\sigma^2}{2}\left(1+\frac{\alpha}{(1-2\logC)^2}[(2\logC-1)(1-2\logC)+2\logC(\logC-1)]\right).
\]
The factor $(2\logC-1)(1-2\logC)+2\logC(\logC-1)=-2\logC^2+2\logC-1$ of the second term of this derivative is a downward facing parabola with a maximum of $-1/2$ at $\logC=1/2$. As this factor is negative for all $\logC$ and $\alpha\le 1$ for $-1/(k-1)\le \rho \le 1$,
\[
\begin{aligned}
\frac{\partial R}{\partial \logC}&\ge \frac{\sigma^2}{2}\left(1+\frac{1}{(1-2\logC)^2}[(2\logC-1)(1-2\logC)+2\logC(\logC-1)]\right)\\
&=\frac{\sigma^2}{2}\frac{(1-2\logC)^2-2\logC^2+2\logC-1}{(1-2\logC)^2}=\frac{\sigma^2}{2}\frac{2\logC(\logC-1)}{(1-2\logC)^2}
\end{aligned}
\]
which is strictly positive whenever $\logC<0.$ Hence, for $\logC<0$, $r_N$ is an increasing function of $\logC.$
}

To understand the effect of $\sigma^2$ on $r_N$ \red{for $\logC<0$, notice that $R$ is linear in $\sigma^2$ with slope
\[
\frac{\logC}{2}\left(1+\alpha\frac{\logC-1}{1-2\logC}\right)=\frac{(1-\alpha)\logC+(\alpha-2)\logC^2}{2-4\logC}
\]
which is negative for $\logC<0$ as $\alpha\le 1$. Hence, $r_N(\XX)$ is a decreasing function of $\sigma^2$ for $\logC<0.$}

\section{Proofs of Theorems~\ref{thm:Xpersistence},\ref{thm:Npersistence1}, and \ref{thm:Npersistence2}}\label{appendixB}

In this Appendix, Theorems~\ref{thm:Xpersistence},\ref{thm:Npersistence1}, and \ref{thm:Npersistence2} are proved using results from \citet{arXiv-18}. To use these theorems, one needs to ensure that the main assumptions (A1-A4) in \citep{arXiv-18} are satisfied. To this end, I assume that the log-fitness function $\log\phi$ is uniformly bounded with respect to the distribution of the $Y_i(t)$ and the population densities remain bounded. Specifically, 
\begin{description}
	\item [A1] \red{Assume $Y_i$ is compactly supported i.e. there exists $a<b$ such that $\P[Y_i\in[a,b]]=1$.} 
	\item [A2] There exists a $K>0$ such that $N(t)$ enters and remains in the compact interval $[0,K]$ with probability one.
\end{description}

\subsection*{Proof of Theorem~\ref{thm:Xpersistence}} Assume $\varepsilon>0$ is sufficiently small, $C<1$, and $\rho<1$. Then Lemma~\ref{lm:ri} implies that for every ergodic, stationary distribution $\XX$ supporting a subset $I\subset\{1,\dots,k\}$ \red{of} alleles,  $r_i(\XX)>0$ for all $i\notin I$. \citet[Theorem 1]{arXiv-18} with $p_1=\dots=p_k=1$, ${\mathcal S}=\Delta$, and ${\mathcal S}_0=\Delta_0$ implies $\Delta_0$-stochastic persistence. 

Next, assume $\varepsilon>0$ is sufficiently small, $C>1$, and $\rho<1$. Then Lemma~\ref{lm:ri} implies that for every ergodic, stationary distribution $\XX$ supporting a subset $I\subset\{1,\dots,k\}$ \red{of} alleles,  $r_i(\XX)<0$ for all $i\notin I$. In particular, for a stationary solution supported on a vertex of the simplex $\Delta$, $r_i(\XX)<0$ for all the strategies not supported by that vertex. Define the set $\Delta^*=\{x\in \Delta: x_i=1$ for some $i\}$ to be the set of vertices of $\Delta$. I will show that this set is \emph{accessible}: for all $\varepsilon>0$ there is $\delta>0$ such that 
\[
\P\left[X_i(t)\ge 1-\varepsilon \mbox{ for some }i \mbox{ and }t\ge 1   \right]\ge \delta
\]
whenever $\prod_i X_i(0)>0.$ Having shown this, \citet[Corollary 2]{arXiv-18} implies that $X(t)$ converges with probability one to one of the vertices as $t\to\infty.$ To show $\Delta^*$ is accessible, consider $X(0)=x$ such that $\prod_i x_i>0.$ Since $\sum_i x_i=1$, there exists some $i$ such that $x_i\ge 1/k$. Without loss of generality (i.e. renaming the indices if needed), I will assume that $x_1\ge 1/k.$ 

As $\rm{Var}[Z_i(t)]=1$ for all $i$ and $\rho<1$, there exist $\delta>0$ and $\eta>0$ such that $\P[Z_1(t)>\max_{i\ge 2}Z_i(t)+\delta]\ge \eta.$  For any $t\ge 1$, define the event ${\mathcal E}(t)=\{Z_1(s)>\max_{i\ge 2}Z_i(s)+\delta$ for $1\le s\le t\}$. By independence in time,  $\P[{\mathcal E}(t)]\ge \eta^t$. Assumption A1 and
\[
\log W_i(X(t),Y(t))=  \varepsilon\sigma \sum_{j=1}^k X_j(t)\frac{Z_i(t)+Z_j(t)}{2}+\rm{O}(\varepsilon^2),
\]
imply that for $i\neq 1$ and $1\le s\le t$
\[
\begin{aligned}
\log \frac{W_1(X(s),Y(s))}{W_i(X(s),Y(s))}=&
\varepsilon \sigma \sum_j X_j(s)\left(\frac{Z_1(s)+Z_j(s)}{2}-\frac{Z_i(s)+Z_j(s)}{2}\right)+O(\varepsilon^2)\\
=&\varepsilon \sigma \frac{Z_1(s)-Z_i(s)}{2}+O(\varepsilon^2).
\end{aligned}
\]
Hence, for $\varepsilon>0$ sufficiently small
\[
\log\red{\frac{W_1(X(s),Y(s))}{W_i(X(s),Y(s))}} \ge \varepsilon\sigma \delta/4 \mbox{ for }1\le s\le t, i\ge 2
\]
 on the event ${\mathcal E}(t)$. Thus, for $\varepsilon>0$ sufficiently small
\[
\log \frac{X_1(t)}{X_i(t)}\ge \log\frac{X_1(0)}{X_i(0)}+ t\sigma \varepsilon\delta/4 \mbox{ for }i\ge 2
\]
on the event ${\mathcal E}(t).$ As $X_1(0)\ge 1/k$, for $\varepsilon>0$ sufficiently small,
\[
X_i(t)\le \exp(-t\sigma \varepsilon\delta/4) X_i(0)/X_1(0)\le \exp(-t\sigma\varepsilon \delta/4) k
\]
on the event ${\mathcal E}(t).$ As the right side can be made arbitrarily small for $t$ sufficiently large and this event occurs with a positive probability $\eta^t$ for any positive initial condition, it follows that $\Delta^*$ is accessible and $\lim_{t\to\infty} X_i(t)=0$ for $i\ge 2$ with probability one.

\subsection*{Proof of Theorems~\ref{thm:Npersistence1} and \ref{thm:Npersistence2}} Assumption A2 implies that the state space for the full model is $M=\Delta \times [0,K]$ and the extinction set is $M_0=\{(X,N)\in M: N\prod_{i=1}^k X_i=0\}$ which corresponds to either the population being extinct (i.e. $N=0$) or one of the alleles missing (i.e. $X_i=0$ for some $i$). 

I only prove Theorem~\ref{thm:Npersistence2} as the proof of Theorem~\ref{thm:Npersistence1} is nearly identical. Assume $C<1$, $\rho<1$, and $\varepsilon>0$ is sufficiently small. Then Lemma~\ref{lm:ri} implies there exists $\eta_1>0$ such that $\sum_ir_i(\XX)>\eta_1$ for every ergodic stationary distribution $\XX$ supporting a subset of the alleles. \red{By assumption A1 and continuity of $\phi$, there exists $\eta_2>0$ such that $r_N(\XX)\le -\eta_2$ all ergodic stationary distributions $\XX$ supporting a subset of alleles i.e. the law of $\XX$ is supported by $\Delta_0$.} Now suppose that
\[
\mu+\frac{\sigma^2\cu}{4}+\frac{\sigma^2(\cu-2)}{4}\frac{
1+2(1-\cu)(\frac{k-1}{k}\rho+\frac{1}{k})}{3-2\cu}>0
\]
Then Lemma~\ref{lm:rN} implies that there exists $\eta_3>0$ such that $r_N(\XX)\ge \eta_3$ for every ergodic stationary distribution supporting all of the alleles i.e. supported on $\Delta\setminus \Delta_0$. Define $p_i=1$ for $i=1,\dots,k$ and $p_N=\eta_1/(2\eta_2)$. Then 
\[
\sum_i p_i r_i (\XX)+p_N r_N(\XX)\ge \frac{\eta_1}{2}\min\{1,\eta_3/\eta_2\}>0
\]
for any ergodic stationary distribution $\XX$ on $M_0$. \citet[Theorem 1]{arXiv-18} with ${\mathcal S}=M$, and ${\mathcal S}_0=M_0$ implies $M_0$-stochastic persistence.

Now suppose that
\[
\mu+\frac{\sigma^2\cu}{4}+\frac{\sigma^2(\cu-2)}{4}\frac{
1+2(1-\cu)(\frac{k-1}{k}\rho+\frac{1}{k})}{3-2\cu}<0.
\]
Then $r_N(\XX)<0$ for every ergodic stationary distribution supporting all of the alleles i.e. $S(\XX)=\{1,2,\dots,k\}.$ For $\varepsilon>0$ sufficiently small, our approximations imply that $r_N(\XX)<0$ for every ergodic distribution supporting any number \red{of }alleles i.e. more alleles always increase the realized per-capita growth rate. Hence, by weak* compactness \red{of Borel probability measures on $\Delta_0\times [0,K]$} and the ergodic decomposition theorem~\citep[see, e.g.,][]{mane-83} \red{which (roughly) states that the law of any stationary distributions can be written as a convex combination of the laws of ergodic stationary distributions}, there exists $\eta>0$ such that $r_N(\XX)\le -\eta$ for any stationary distribution $\XX$ (including non-ergodic ones). Using a standard argument (see, e.g., the first paragraph of the proof of Proposition 2 in \citet[Section 7.2]{arXiv-18}), it follows that 
\[
\limsup_{t\to\infty} \frac{1}{t}\sum_{s=0}^{t-1}\log W(X(s),Y(s)) \le -\eta \mbox{ with probability one}
\] 
for any initial conditions $X(0)$. As the density dependent reduction term $\red{f}(N)$ is a decreasing function with $\red{f}(0)=1$, it follows that  
\begin{eqnarray*}
\limsup_{t\to\infty}\frac{1}{t}\log N(t)&=&\limsup_{t\to\infty} \frac{1}{t}\sum_{s=0}^{t-1}\log W(X(s),Y(s))\red{f}(N(s))\\
 &\le &\limsup_{t\to\infty}
\frac{1}{t}\sum_{s=0}^{t-1}\log W(X(s),Y(s))\\
& \le &-\eta \mbox{ with probability one}
\end{eqnarray*}
whenever $N(0)>0.$ This completes the proof of Theorem~\ref{thm:Npersistence2}.
\end{document}